\documentclass[aps,prb,10pt,twocolumn,superscriptaddress]{revtex4-2}

\usepackage{graphicx}
\usepackage{amssymb,amsmath}
\usepackage{color}

\newcommand{\bfr}{{\bf r}}
\newcommand{\bfx}{{\bf x}}
\newcommand{\bfR}{{\bf R}}
\newcommand{\bfk}{{\bf k}}
\newcommand{\bfA}{{\bf A}}
\newcommand{\bfE}{{\bf E}}

\begin{document}

\title{Time-resolved exciton wave functions from time-dependent density-functional theory}

\author{Jared R. Williams}
\affiliation{Department of Physics and Astronomy, University of Missouri, Columbia, Missouri 65211, USA}

\author{Nicolas Tancogne-Dejean}
\affiliation{Max Planck Institute for the Structure and Dynamics of Matter, 22761 Hamburg, Germany}

\author{Carsten A. Ullrich}
\affiliation{Department of Physics and Astronomy, University of Missouri, Columbia, Missouri 65211, USA}

\date{\today }


\begin{abstract}
Time-dependent density-functional theory (TDDFT) is a computationally efficient first-principles approach
for calculating optical spectra in insulators and semiconductors, including excitonic effects.
We show how exciton wave functions can be obtained from TDDFT via the Kohn-Sham transition density matrix, both in the
frequency-dependent linear-response regime and in real-time propagation. The method is illustrated using one-dimensional model solids. In particular, we show that our approach provides insight into the formation and dissociation of excitons in real time. This opens the door to time-resolved studies of exciton dynamics in materials
by means of real-time TDDFT.
\end{abstract}

\maketitle

\section{Introduction}\label{sec:I}

According to the common textbook definition \cite{Kittel}, an exciton is a bound electron-hole pair which
is created in an optical excitation of an insulator or semiconductor across the band gap. Within the so-called Wannier exciton model \cite{Haug},
the electron and the hole attract each other via the dielectrically screened Coulomb interaction; the exciton wave function follows from
a hydrogenic Schr\"odinger equation for a particle of reduced mass $m_r = m_em_h/(m_e+m_h)$, where $m_e$ and $m_h$ are the electron and
hole effective mass, respectively. In three dimensions, the Wannier model gives a qualitatively useful picture of excitons in solids, but it is too simplistic
for most applications, especially when one is interested in the dynamics of the exciton induced by some external perturbation.

The Bethe-Salpeter equation (BSE), usually coupled with the GW method for the electronic band structure, is a well-established first-principles approach
for calculating optical spectra including excitonic features \cite{Onida2002,Bechstedt2015,Reining2016}.
More recently, time-dependent density-functional theory (TDDFT) \cite{Ullrich2012} has made significant progress as an alternative approach
for the optical properties of semiconductors and insulators \cite{Ullrich2015,Byun2017b,Byun2020}, with accuracies close to the BSE,
but at a fraction of its cost \cite{Sun2020,Sun2020a}.

The BSE can be formulated as an eigenvalue problem, giving rise to the excitation spectrum; the eigenvectors can be used to construct a two-body object,
the exciton wave function $\Psi(\bfr_e,\bfr_h)$, where $\bfr_e$ and $\bfr_h$ are the positions of the electron and the hole, respectively.
There are many examples in the literature where BSE exciton wave functions are studied for various materials
\cite{Rohlfing2000,Puschnig2002,Hummer2004,Hummer2005,Laskowski2005,Galamic2005,Huang2013,Erhart2014,Tran2014,Sharifzadeh2018,Lewis2019}. However,
this kind of analysis is limited to the frequency-dependent linear-response regime; the standard BSE does not
tell us how excitonic wave functions have been created, nor how they behave
under more general time-dependent perturbations such as sudden switching or short laser pulses, especially if the resulting dynamics is ultrafast or goes beyond
the linear regime. There exists an explicitly time-dependent version of the BSE \cite{Rabani2015}, but to our knowledge it has not been used to
obtain exciton wave functions.

In this paper, we show how exciton wave functions can be obtained from TDDFT, both in the frequency-dependent linear-response and in the real-time propagation regime.
This makes it possible to investigate from first principles how excitons evolve under non-steady-state conditions.
An alternative approach for real-time exciton dynamics is based on nonequilibrium Green's functions, which involves the solution of the
time-dependent Kadanoff-Baym equation \cite{Stefanucci2013}. Such calculations have recently been implemented for molecules and solids \cite{Perfetto2015,Perfetto2016,Perfetto2018,Sangalli2018,Perfetto2019},
but the computational effort for realistic materials is significantly higher than that of TDDFT.

In order to illustrate the dynamics of the exciton wave function,
we will use one-dimensional (1D) models of a solid, for which the entire two-body wave function $\Psi(\bfr_e,\bfr_h)$ can be plotted as a two-dimensional map.
For further analysis, especially to illustrate the time evolution, we will also use representations of reduced dimensionality, where we fix the position of the
hole and plot the distribution of the electron
around it (alternatively, one can also fix the center of mass coordinate of the exciton and plot the relative electron-hole coordinate). By fixing the position of the hole, or of the center of mass, we can easily visualize the dynamics of the exciton wave function, a task that becomes much more complicated in higher dimensions.

This paper is organized as follows. In Section \ref{sec:II} we discuss the theoretical background, introducing the frequency- and time-dependent
transition density matrix and its representation in periodic solids. We also discuss issues related to gauge invariance.
In Section \ref{sec:III} we present results for 1D model solids, illustrating the exciton wave function in frequency-dependent and real-time representations.
We will discuss explicit examples to showcase the capabilities of the approach, namely,
the visualization of localized and charge-transfer excitons, the onset of nonlinear effects under increasing excitation strength,
and exciton dissociation under the influence of static electric fields.
Conclusions are given in Section \ref{sec:IV}.

Atomic units (a.u.), with $\hbar=e=m=4\pi\epsilon_0$=1, will be used throughout. Explicit values of
physical quantities such as lengths, energies, or electric field strengths will be given as dimensionless numbers; it is
understood that they are measured in a.u.

\section{Theoretical background}\label{sec:II}

\subsection{TDM of the $n$th excitation}

The transition density matrix (TDM) between a many-body ground state $\Psi_0$ and the $n$th excited state $\Psi_n$ is defined as
\cite{McWeeny1960,Furche2001}
\begin{equation}
\Gamma^{(n)}(\bfr,\bfr') = \langle \Psi_n | \hat \rho(\bfr,\bfr') | \Psi_0 \rangle,
\end{equation}
where $\hat\rho(\bfr,\bfr')$ is the reduced one-particle density matrix operator.

The TDM can be approximately obtained from Kohn-Sham- or quasi-particle-based theories such as TDDFT, generalized (hybrid) TDDFT, time-dependent Hartree-Fock, or GW/BSE.
In each case, the first step is to calculate the spectrum of excitation energies $\Omega_n$, which involves solving a non-Hermitian eigenvalue
equation in one-particle transition space of the form
\begin{equation}\label{Casida}
\left( \begin{array}{cc} {\bf A} & {\bf B} \\ {\bf B^*} & {\bf A^*}\end{array}\right)
\left(\begin{array}{c} {\bf X}^{(n)} \\ {\bf Y}^{(n)} \end{array}\right) = \Omega_n
\left( \begin{array}{cc} -{\bf 1} & {\bf 0} \\ {\bf 0} & {\bf 1}\end{array}\right)
\left(\begin{array}{c} {\bf X}^{(n)} \\ {\bf Y}^{(n)} \end{array}\right).
\end{equation}
In TDDFT, the elements of the matrix $\bf A$ are given by
\begin{equation}
A_{ia,i'a'} = (\varepsilon_a - \varepsilon_i)\delta_{ii'}\delta_{aa'} + B_{ia,i'a'} \:,
\end{equation}
where $i,i'$ and $a,a'$ refer to occupied and unoccupied single-particle levels, respectively, and $\varepsilon_a - \varepsilon_i$ are the
single-particle excitation energies. The matrix $\bf B$ is defined as
\begin{equation}
B_{ia,i'a'} = \!\int \! d\bfr \!\! \int \! d\bfr'\varphi_i^*(\bfr)\varphi_a(\bfr) f_{\rm Hxc}(\bfr,\bfr',\omega)\varphi_{i'}(\bfr')\varphi_{a'}^*(\bfr'),
\end{equation}
where $f_{\rm Hxc}$ is the (formally frequency-dependent) Hartree-exchange-correlation kernel. In TDDFT, Eq. (\ref{Casida}) is known
as the Casida equation \cite{Casida1995}.

In hybrid TDDFT, time-dependent Hartree-Fock, or in GW/BSE, the coupling matrices $\bf A$ and $\bf B$ are defined in a similar manner as in Kohn-Sham TDDFT,
involving double integrals of single-particle orbitals with the bare or screened Coulomb interaction. Explicit expressions
can be found in the literature \cite{Dreuw2005,Ullrich2012,Sun2020}.

Using the occupied and unoccupied single-particle orbitals and the eigenvectors $({\bf X}^{(n)},{\bf Y}^{(n)})$, the TDM can be constructed as
\begin{equation}\label{TDMns}
\Gamma^{(n)}_s(\bfr,\bfr') = \sum_{ia}[\varphi_i(\bfr)\varphi_a^*(\bfr')X_{ia}^{(n)} + \varphi_i^*(\bfr')\varphi_a(\bfr)Y_{ia}^{(n)}].
\end{equation}
Here, the subscript $s$ stands for ``single-particle''.

In TDDFT, the diagonal of the single-particle TDM is the density response associated with
the $n$th excitation,  which is given in principle exactly:
$\Gamma^{(n)}_s(\bfr,\bfr) = \Gamma^{(n)}(\bfr,\bfr) = \delta n(\bfr,\Omega_n)$. However, the nondiagonal elements (where $\bfr\ne\bfr'$)
are not guaranteed to be exact. Nevertheless, $\Gamma^{(n)}_s(\bfr,\bfr')$ has been widely used to analyze electronic excitations \cite{Tretiak2002}.

\subsection{Time-dependent TDM}
Let us now assume that the system starts from the ground state at time $t=0$ and then evolves for $t>0$  under the influence of
time-dependent scalar or vector potentials. We define the time-dependent TDM as the difference
between the time-dependent and the ground-state one-body density matrices:
\begin{equation}\label{6}
\Gamma(\bfr,\bfr',t) = \langle \Psi(t) | \hat \rho(\bfr,\bfr')| \Psi(t)\rangle -
\langle \Psi_0 | \hat \rho(\bfr,\bfr')| \Psi_0\rangle,
\end{equation}
where $\Psi(t)$ is the time-dependent many-body wave function which evolves from the initial state $\Psi_0$.
The time-dependent density matrix $\langle \Psi(t) | \hat \rho(\bfr,\bfr')| \Psi(t)\rangle$ is commonly used for describing
nonequilibrium electronic processes such as transient absorption spectroscopy \cite{Perfetto2015}. Defining the time-dependent
TDM as in Eq. (\ref{6}) allows us to visualize the dynamical changes induced by the external perturbation, and is
consistent with the linear-response TDM, as we will show below.

The time-dependent wave function can be written as
\begin{equation}
\Psi(t) = \Psi_0 e^{-iE_0 t} + \delta \Psi(t) \:,
\end{equation}
and to first order in $\delta \Psi(t)$ the time-dependent TDM becomes \cite{Li2011}
\begin{eqnarray}
\delta \Gamma(\bfr,\bfr',t) &=& \langle \delta \Psi(t) | \hat \rho(\bfr,\bfr')| \Psi_0\rangle e^{-iE_0 t}  \nonumber\\
&+&
e^{iE_0 t}\langle  \Psi_0   | \hat \rho(\bfr,\bfr')|\delta \Psi(t)\rangle.
\end{eqnarray}

In real-time TDDFT, the time-dependent TDM is given by
\begin{equation}\label{TDDMR}
\Gamma_s(\bfr,\bfr',t) = \sum_i^{\rm occ}\left[\varphi_i(\bfr,t)\varphi_i^*(\bfr',t) - \varphi_i(\bfr)\varphi_i^*(\bfr')\right],
\end{equation}
where the  $\varphi_i(\bfr,t)$ are the time-dependent Kohn-Sham orbitals which evolve from the $i$th occupied Kohn-Sham orbitals $\varphi_i(\bfr)$ in the ground state at time $t=0$.

Writing the time-dependent Kohn-Sham orbitals as
\begin{equation}
\varphi_i(\bfr,t) = \varphi_i(\bfr)e^{-i\varepsilon_i t} + \delta\varphi_i(\bfr,t) \:,
\end{equation}
we obtain, to first order,
\begin{eqnarray}\label{TDTDM}
\delta\Gamma_s(\bfr,\bfr',t) &=& \sum_i^{\rm occ}\Big[\varphi_i(\bfr)e^{-i \varepsilon_i t}\delta\varphi_i^*(\bfr',t) \nonumber\\
&&{}
+ \delta\varphi_i(\bfr,t)\varphi_i^*(\bfr')e^{i\varepsilon_i t}\Big].
\end{eqnarray}
It is straightforward to establish a correspondence with the TDM of the $n$th excitation, Eq. (\ref{TDMns}),
by expressing the time-evolved states in the basis of the ground-state Kohn-Sham orbitals. This leads to
\begin{equation}
\delta\varphi_i(\bfr,t) = \sum_k C_{ik}(t)e^{-i\varepsilon_k t}\varphi_k(\bfr)\:.
\end{equation}
Inserting this into Eq. (\ref{TDTDM}), and using standard first-order time-dependent perturbation theory, it follows that
\begin{eqnarray}\label{TDTDM4}
\delta\Gamma_s(\bfr,\bfr',t) &=&
\sum_i^{\rm occ}\sum_a^{\rm unocc}\Big[C^*_{ia}(t)e^{i(\varepsilon_a-\varepsilon_i) t}\varphi_i(\bfr)\varphi^*_a(\bfr') \nonumber\\
&&{}
+ C_{ia}(t)e^{i(\varepsilon_i-\varepsilon_a) t}\varphi_a(\bfr)\varphi_i^*(\bfr')\Big].
\end{eqnarray}
Assuming that the system is in an eigenmode associated with the $n$th excitation, one is then able to identify
the Fourier transforms of the coefficients $C_{ia}(t)e^{-i(\varepsilon_a-\varepsilon_i) t}$ and
$C_{ia}^*(t)e^{i(\varepsilon_a-\varepsilon_i) t}$ with $X_{ia}^{(n)}$ and $Y_{ia}^{(n)}$, respectively \cite{Li2016}.
This allows one to obtain $X_{ia}^{(n)}$ and $Y_{ia}^{(n)}$ without solving the Casida equation, by using time propagation following a small kick at the initial time.

\subsection{Exciton wave functions from TDDFT}

We now proceed to make a more direct connection between the exciton wave function and the single-particle TDM obtained from TDDFT.
The TDM has been identified with the exciton wave function in large molecular systems \cite{Bappler2014}.

In the case of periodic solids, the Casida equation (\ref{Casida}) can be generalized in a rather straightforward manner \cite{Sun2020}.
In the following, we will specifically consider solids with a gap, i.e., semiconductors or insulators, in which excitonic effects can be observed.
In solids, Eq. (\ref{TDMns}) becomes
\begin{equation}\label{TDMk1}
\Gamma^{(n)}_s(\bfr,\bfr') = \sum_{vc\bfk}[\varphi_{v\bfk}(\bfr)\varphi_{c\bfk}^*(\bfr')X_{vc\bfk}^{(n)} + \varphi_{v\bfk}^*(\bfr')\varphi_{c\bfk}(\bfr)Y_{vc\bfk}^{(n)}].
\end{equation}
Here, $v$ and $c$ are valence- and conduction band indices, $\bfk$ is a wave vector within the first Brillouin zone, and $\bfr$ and $\bfr'$ are
arbitrary positions within the periodic solid, not restricted to one unit cell. This reflects the fact that the exciton is an extended object that lives in the entire
periodic crystal; the total size of the crystal is determined by the number of $\bfk$-points used to sample the Brillouin zone.
However, Eq. (\ref{TDMk1}) can be brought into an alternative, more convenient, form by defining $\bfr = \bfx + \bfR$, where $\bfx$ is within the Wigner-Seitz
unit cell, and $\bfR$ is a direct lattice vector. Making use of Bloch's theorem for the single-particle wave functions, we find
\begin{eqnarray}\label{TDMk2}
\Gamma^{(n)}_s(\bfr,\bfr') &=& \sum_{vc\bfk}\Big[\varphi_{v\bfk}(\bfx)\varphi_{c\bfk}^*(\bfx')X_{vc\bfk}^{(n)} \nonumber\\
&+&
\varphi_{v\bfk}^*(\bfx')\varphi_{c\bfk}(\bfx)Y_{vc\bfk}^{(n)}\Big]e^{i\bfk\cdot(\bfR-\bfR')}.
\end{eqnarray}
From this expression, it is clear that in order to construct the TDM in the entire periodic crystal, only input from within a unit cell is required.
If the $n$th excitation has excitonic character, then Eq. (\ref{TDMk2}) gives the exciton wave function.
This will be illustrated in Section \ref{sec:III} for model insulators.

In a similar manner, the time-dependent TDM (\ref{TDDMR}) can be formulated for periodic solids:
\begin{eqnarray}\label{TDDMRp}
\Gamma_s(\bfr,\bfr',t) &=& \sum_{v\bfk}\Big[\varphi_{v\bfk}(\bfx,t)\varphi_{v\bfk}^*(\bfx',t)\nonumber\\
&-&
\varphi_{v\bfk}(\bfx)\varphi_{v\bfk}^*(\bfx')\Big]e^{i\bfk\cdot(\bfR-\bfR')}
\end{eqnarray}
and likewise for the first-order expression (\ref{TDTDM}),
\begin{eqnarray}\label{TDTDMp}
\delta\Gamma_s(\bfr,\bfr',t) &=& \sum_{v\bfk}\Big[\varphi_{v\bfk}(\bfx)e^{-i \varepsilon_{v\bfk} t}\delta\varphi_{v\bfk}^*(\bfx',t) \nonumber\\
&+&
\delta\varphi_{v\bfk}(\bfx,t)\varphi_{v\bfk}^*(\bfx')e^{i\varepsilon_{v\bfk} t}\Big]e^{i\bfk\cdot(\bfR-\bfR')}.
\end{eqnarray}

In practice, the time-dependent TDM is constructed using orbitals obtained from numerical solutions of the time-dependent Kohn-Sham equations;
the full expressions (\ref{TDDMR}) and (\ref{TDDMRp}) for $\Gamma_s$ are then to be preferred over the linearized expressions (\ref{TDTDM}) and (\ref{TDTDMp}).
The main reason is that the phase factors $e^{-i \varepsilon_i t}$ and $e^{\pm i\varepsilon_{v\bfk} t}$, respectively, depend on the Kohn-Sham
single-particle eigenvalues, which in practice are determined to within some numerical error; this introduces
numerical inaccuracies which will accumulate over time.

\subsection{Gauge dependence of the time-dependent TDM}

We first consider the time-dependent Kohn-Sham equation in the length gauge:
\begin{eqnarray}\label{KS}
\lefteqn{
i \frac{\partial}{\partial t}\varphi_j(\bfr,t) =}\\
&& \left[-\frac{\nabla^2}{2} +  V_0(\bfr) +  V_1(\bfr,t) + V_{\rm Hxc}(\bfr,t) \right]\varphi_j(\bfr,t) \:. \nonumber
\end{eqnarray}
We assume that the system is initially in the ground state associated with the static external potential $V_0(\bfr)$; at time $t=0$,
a time-dependent scalar potential $V_1(\bfr,t)$ is switched on and the system is driven out of the  ground state.
$V_{\rm Hxc}(\bfr,t)$ is the sum of the time-dependent Hartree and exchange-correlation potentials.

To describe optical processes in materials, it is convenient to transform
the time-dependent Kohn-Sham equation (\ref{KS}) into the velocity gauge
\cite{Bertsch2000,Yabana2006,Yabana2012,Yamada2019,Krieger2015,Tancogne2017a,Tancogne2017b,Pemmaraju2018,Gabay2020}:
\begin{eqnarray}\label{KSA}
\lefteqn{
i \frac{\partial}{\partial t}\tilde \varphi_j(\bfr,t) =}\\
&& \left[\frac{1}{2}\left(\frac{\nabla}{i} + \bfA_1(\bfr,t)\right)^2 +  V_0(\bfr) + V_{\rm Hxc}(\bfr,t) \right]\tilde\varphi_j(\bfr,t) \:. \nonumber
\end{eqnarray}
The vector potential $\bfA_1(\bfr,t)$ is given by
\begin{equation}\label{gaugetrafo}
\bfA_1(\bfr,t) = -\nabla\int_{0}^t V_1(\bfr,t')dt' \:,
\end{equation}
which follows from the relation $\partial \bfA_1(\bfr,t)/\partial t = \bfE_1(\bfr,t)$ between the vector potential and the
electric field $\bfE_1$ associated with the scalar potential $V_1$.

The Kohn-Sham orbitals in the velocity gauge, $\tilde\varphi_j(\bfr,t)$, are related to the orbitals in the length gauge,
$\varphi_j(\bfr,t)$, as follows:
\begin{equation}\label{phigauge}
\tilde\varphi_j(\bfr,t) = e^{-i\Lambda(\bfr,t)}\varphi_j(\bfr,t) \:,
\end{equation}
where $\partial \Lambda(\bfr,t)/\partial t = -V_1(\bfr,t)$.

An important example is that of a linearly polarized electromagnetic wave in dipole approximation. The perturbing potential then has the form
$V_1(\bfr,t) = -\bfE \cdot \bfr f(t)$, where $\bfE$ is the constant uniform electric field amplitude, and $f(t)$ is a purely time-dependent
function describing, for instance, a short kick or a short pulse. The associated gauge function and vector potential are then $\Lambda(\bfr,t) = \bfE \cdot \bfr F(t)$ and
$\bfA_1(\bfr,t) = \bfE F(t)$, respectively, where $F(t) = \int_0^t f(t') dt'$.

The time-dependent TDM is not invariant under electromagnetic gauge transformations. If we start from $\Gamma_s(\bfr,\bfr',t)$ in Eq. (\ref{TDDMR})
evaluated with length-gauge wave functions, we obtain, using the above gauge transformation, the TDM as
\begin{eqnarray}\label{TDDMRgauge}
\Gamma_s(\bfr,\bfr',t) &=& \sum_i^{\rm occ}\Big[\tilde\varphi_i(\bfr,t)\tilde\varphi_i^*(\bfr',t) e^{i(\Lambda(\bfr,t)-\Lambda(\bfr',t))}
\nonumber\\
&&{}
 - \varphi_i(\bfr)\varphi_i^*(\bfr')\Big].
\end{eqnarray}
This is clearly different from the time-dependent TDM directly constructed from the wave functions obtained in the velocity gauge,
\begin{equation}\label{TDDMRvel}
\tilde\Gamma_s(\bfr,\bfr',t) = \sum_i^{\rm occ}\left[\tilde\varphi_i(\bfr,t)\tilde\varphi_i^*(\bfr',t) - \varphi_i(\bfr)\varphi_i^*(\bfr')\right].
\end{equation}
As we will see in the next section, the lack of gauge invariance can sometimes be mitigated by a suitable choice of $A_1(\bfr,t)$.
However, we found in all examples of Sec. \ref{sec:IIIB} that the differences between $\Gamma_s(\bfr,\bfr',t)$ and $\tilde\Gamma_s(\bfr,\bfr',t)$ were rather minor.
In the following, we therefore only consider $\tilde\Gamma_s(\bfr,\bfr',t)$, as defined in Eq. (\ref{TDDMRvel}).

An important point to note is that the gauge transformation in Eqs. (\ref{KSA}) and (\ref{gaugetrafo}) has only been applied to the external time-dependent potential.
As we will discuss in more detail below, this is the proper thing to do for 1D systems. However, in 2D and 3D there may be long-range contributions
to the time-dependent exchange-correlation potential, which must be gauge transformed into a vector potential as well.

\begin{figure}
\includegraphics[width=\linewidth]{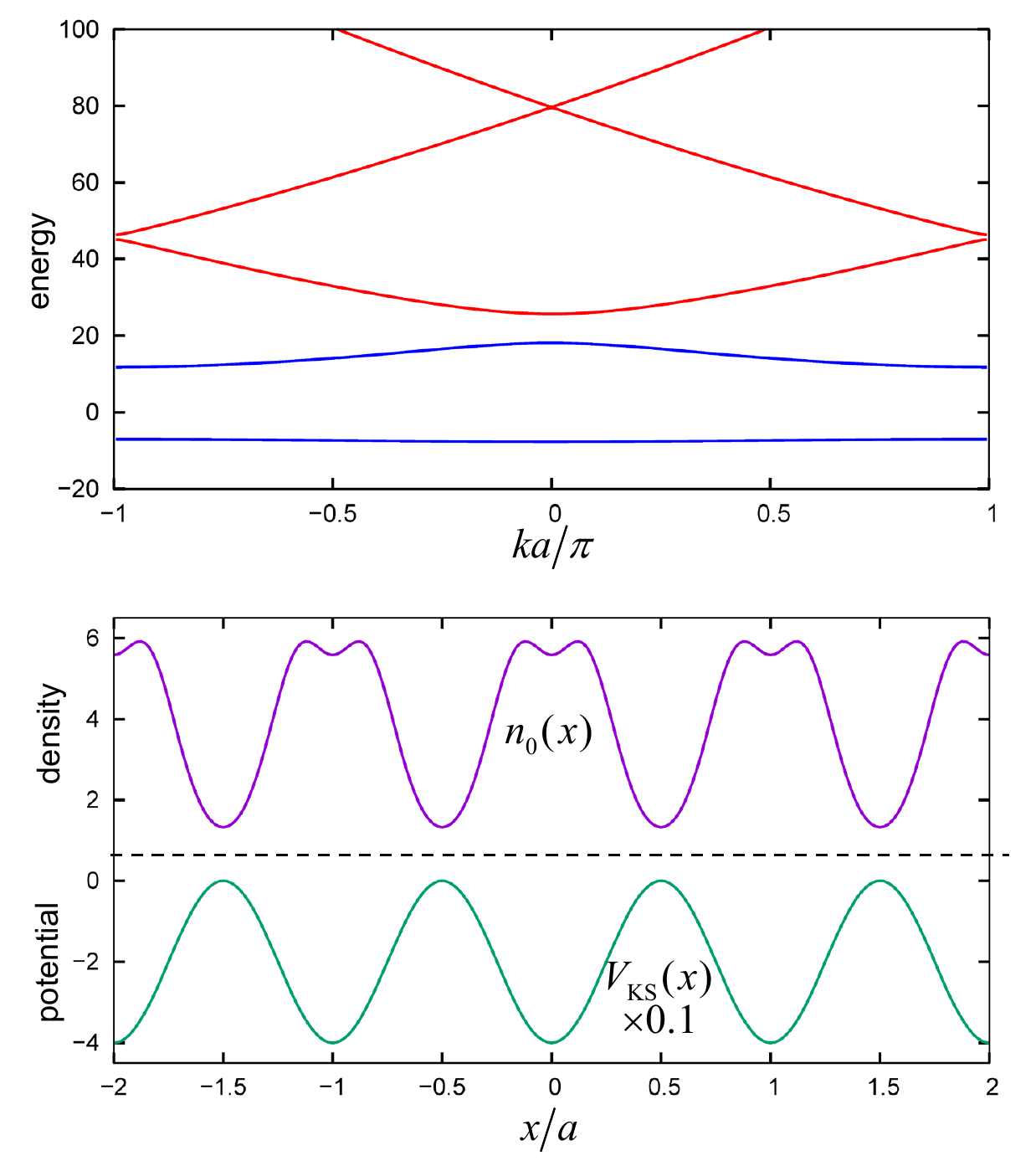}\centering
\caption{Bottom: Kohn-Sham potential $V_{\rm KS}(x)$ (scaled by 0.1) and ground-state density $n_0(x)$ of a model 1D insulator
with four electrons per unit cell. Top: associated band structure (occupied bands in blue, empty bands in red).} \label{fig1}
\end{figure}

\section{Results for model solids}\label{sec:III}

\subsection{Time-independent description of excitons}

\subsubsection{Band structure of the 1D model solid}

We will now illustrate the frequency- and time-dependent TDM using 1D model solids \cite{Yang2012,Johnston2020}.
We first calculate the electron band structure in a 1D periodic Kohn-Sham potential $V_{\rm KS}(x) = V_0(x) + V_{\rm Hxc}(x)$ with cosine shape:
\begin{equation}\label{VKS_cosine}
V_{\rm KS}(x) = -A \cos\left(\frac{2\pi x}{a}\right).
\end{equation}
Here, $a$ is the lattice constant, and $A$ is the amplitude of the potential. Separate knowledge of $V_0(x)$ and $V_{\rm Hxc}(x)$ is not needed here.
In the following, we choose $A=20$  and $a=1$, and
we consider the case where the two lowest bands are occupied, i.e., there are four electrons per unit cell.
The ground-state density $n_0(x)$ and the associated $V_{\rm KS}(x)$ are shown in the bottom panel of Fig. \ref{fig1}.

The top panel of Fig. \ref{fig1} presents the electronic band structure in the first Brillouin zone; occupied valence bands are shown in blue, empty conduction bands in red.
There is a direct band gap of size $E_g = 7.56$ at the $\Gamma$-point (where $k=0$). Here and in the following, the calculations were done using
a straightforward plane-wave expansion with 200 $k$-points in the Brillouin zone and 7 reciprocal lattice vectors. Later, in Section \ref{sec:IIIA3},
we will consider a 1D solid with defects, using a supercell  with 15 $k$-points in the Brillouin zone.

\subsubsection{Excitons from linear-response TDDFT}

Next, we use TDDFT in the frequency-dependent linear-response formalism to calculate the excitation energies.
Specifically, we solve the Casida equation, Eq. (\ref{Casida}), for our 1D solid, including the two occupied valence bands and three unoccupied conduction bands.
We use the following xc kernel:
\begin{equation}\label{fxc_x}
f_{\rm xc}^{\rm LRC}(x,x') = -\frac{\alpha}{\sqrt{(x-x')^2 + \gamma^2}} \:.
\end{equation}
This xc kernel is the 1D version of the so-called long-range corrected (LRC) kernel \cite{Reining2002,Botti2004,Byun2017b,Sun2020},
featuring two adjustable parameters, $\alpha$ and $\gamma$. Here, $\gamma$ defines the 1D soft Coulomb potential; in the following we choose $\gamma=0.1$.
The strength $\alpha$ of the LRC kernel determines the exciton binding energy.

\begin{figure}
\includegraphics[width=\linewidth]{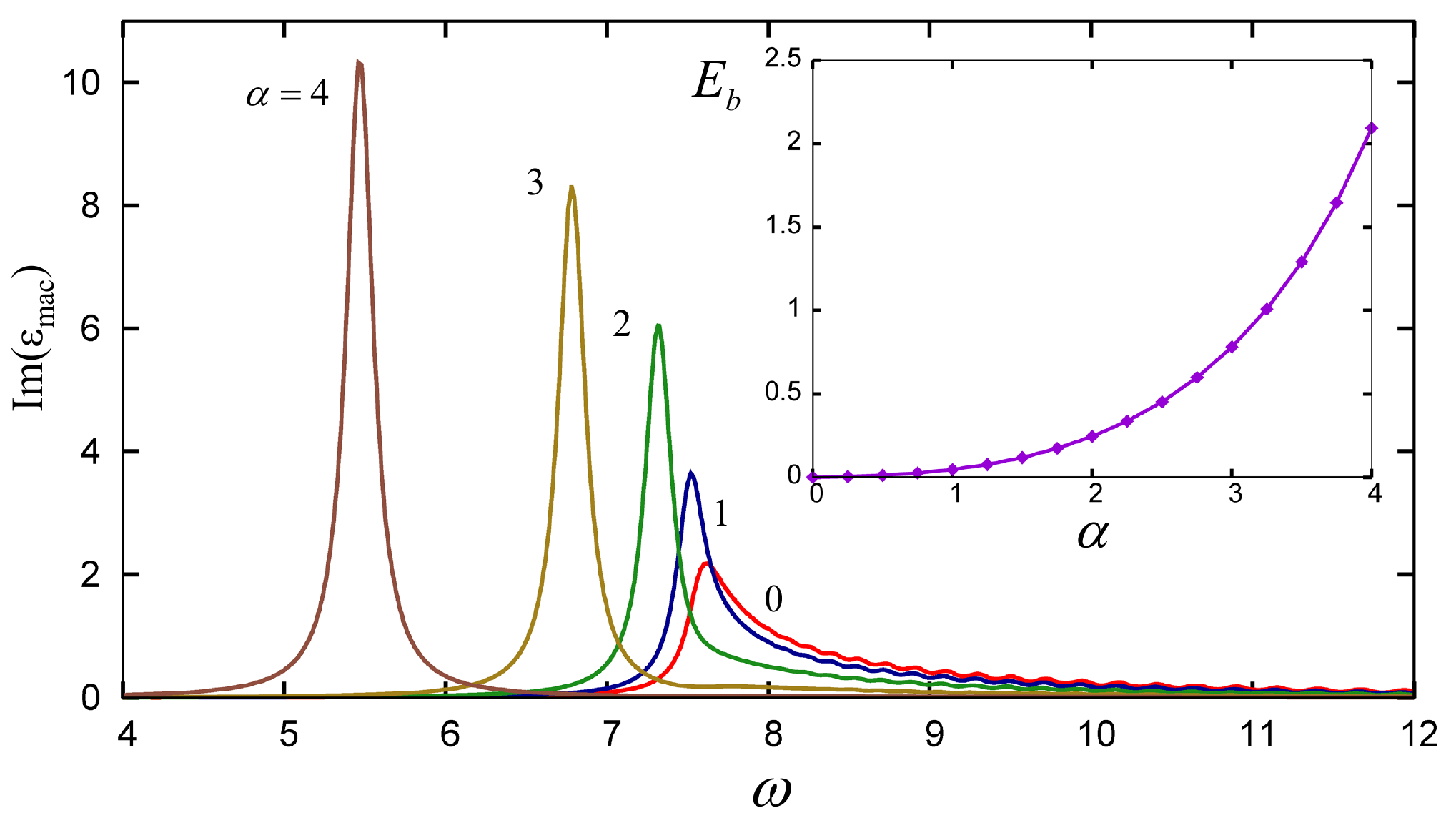}\centering
\caption{Imaginary part of the macroscopic dielectric function of the 1D solid, for various values of the LRC parameter $\alpha$, as indicated.
Inset: exciton binding energy $E_b$ versus  $\alpha$.} \label{fig2}
\end{figure}

In reciprocal space, the xc kernel is given by
\begin{equation}\label{fxc_G}
f^{\rm LRC}_{{\rm xc},GG'}(q) = -2\alpha K_0(\gamma|q+G|)\delta_{GG'} \:,
\end{equation}
where $K_0$ denotes a modified Bessel function of the second kind, $q$ is a wave vector in the first Brillouin zone, and $G,G'$ are reciprocal lattice vectors
of the 1D lattice. Since $\lim_{q\to 0}K_0(\gamma|q|)=\ln (\gamma|q|)$, the head of the 1D LRC kernel diverges logarithmically rather than as $1/q^2$.
Therefore, only the body of $f^{\rm LRC}_{{\rm xc},GG'}$ contributes to the excitonic binding. In other words, the 1D LRC kernel
creates excitons via local-field effects, in contrast with LRC kernels in three dimensions, where the head is dominant \cite{Yang2012,Byun2020}.

\begin{figure}[t]
\includegraphics[width=\linewidth]{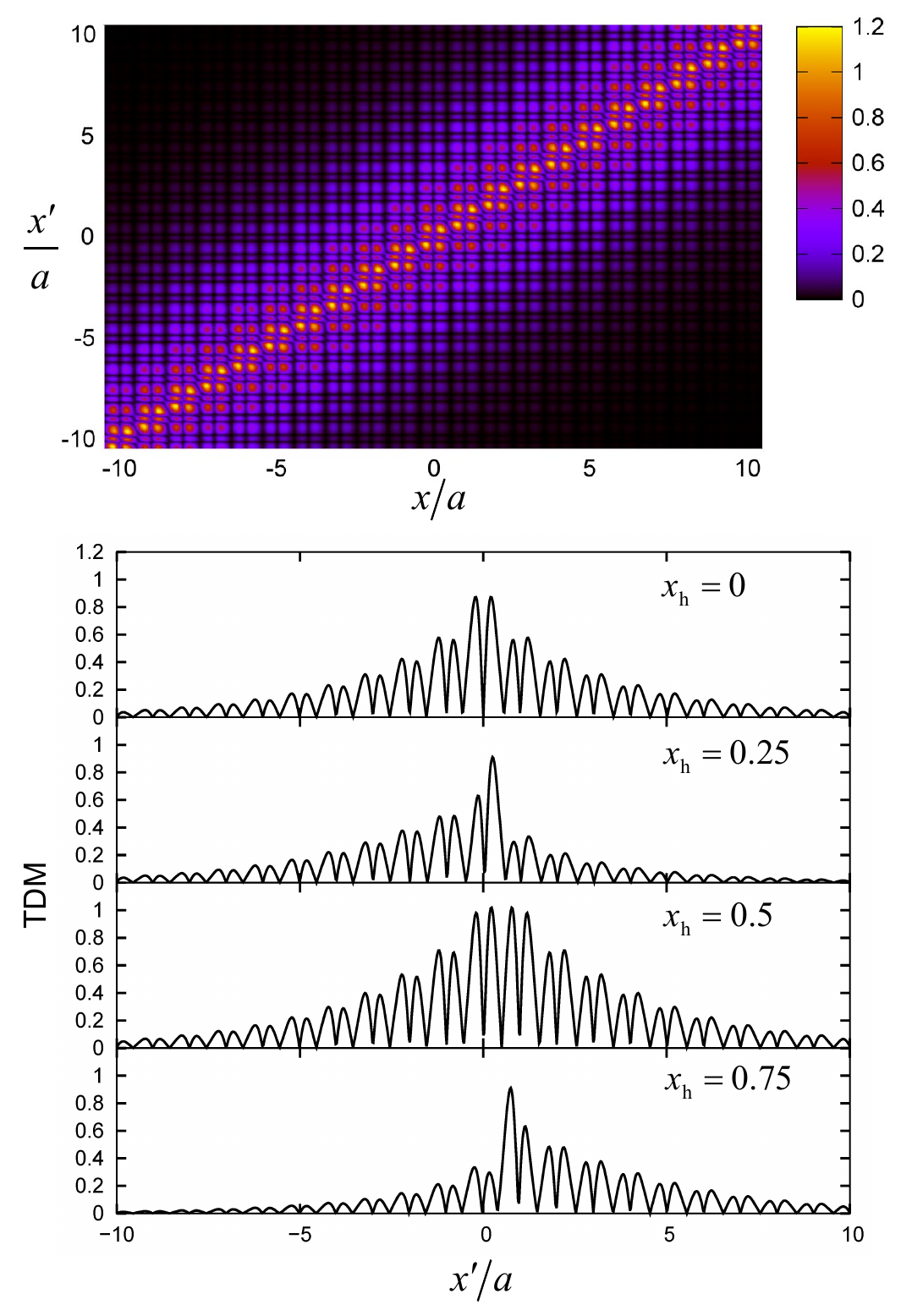}\centering
\caption{Top: frequency-dependent TDM $|\Gamma_s(x,x')|$ of the bound exciton for $\alpha=3$. Bottom: electron distribution $|\Gamma_s(x_{\rm h},x')|$
for various reference positions $x_{\rm h}$ of the hole.
} \label{fig3}
\end{figure}

From the eigenvectors of the Casida equation, the macroscopic dielectric function $\epsilon_{\rm mac}(\omega)$ can be constructed \cite{Byun2020}
\footnote{The logarithmic singularity of the soft-Coulomb interaction makes it necessary to evaluate the 1D macroscopic dielectric function
at a small but finite value of $q$ \cite{Yang2012}. Here, we choose $q=0.01$.}.
Figure \ref{fig2} shows the imaginary part of $\epsilon_{\rm mac}(\omega)$ for our 1D solid, for various values of $\alpha$.
Strong excitonic peaks are seen to develop as $\alpha \gtrsim 1$.
The exciton binding energy $E_b$, shown in the inset, strongly increases with $\alpha$.

Figure \ref{fig3} shows the frequency-dependent TDM of the bound exciton for $\alpha=3$ (corresponding to the peak at $\omega = 6.79$
in Fig. \ref{fig2}, with $E_b=0.78$). The top panel presents the absolute value $|\Gamma_s(x,x')|$, where $x$ and $x'$ cover a range of 21 unit cells.
Clearly, the TDM is diagonally dominated, as one would expect for the wave function of a bound exciton, for which the electrons and holes are held
together by the Coulomb interaction. The bottom panels of Fig. \ref{fig3} show the electron distribution $|\Gamma_s(x_{\rm h},x')|$ for various
reference positions $x_{\rm h}$ of the hole in the central unit cell. Each of the four profiles is a vertical cut through the TDM of the top panel.

\begin{figure}[t]
\includegraphics[width=\linewidth]{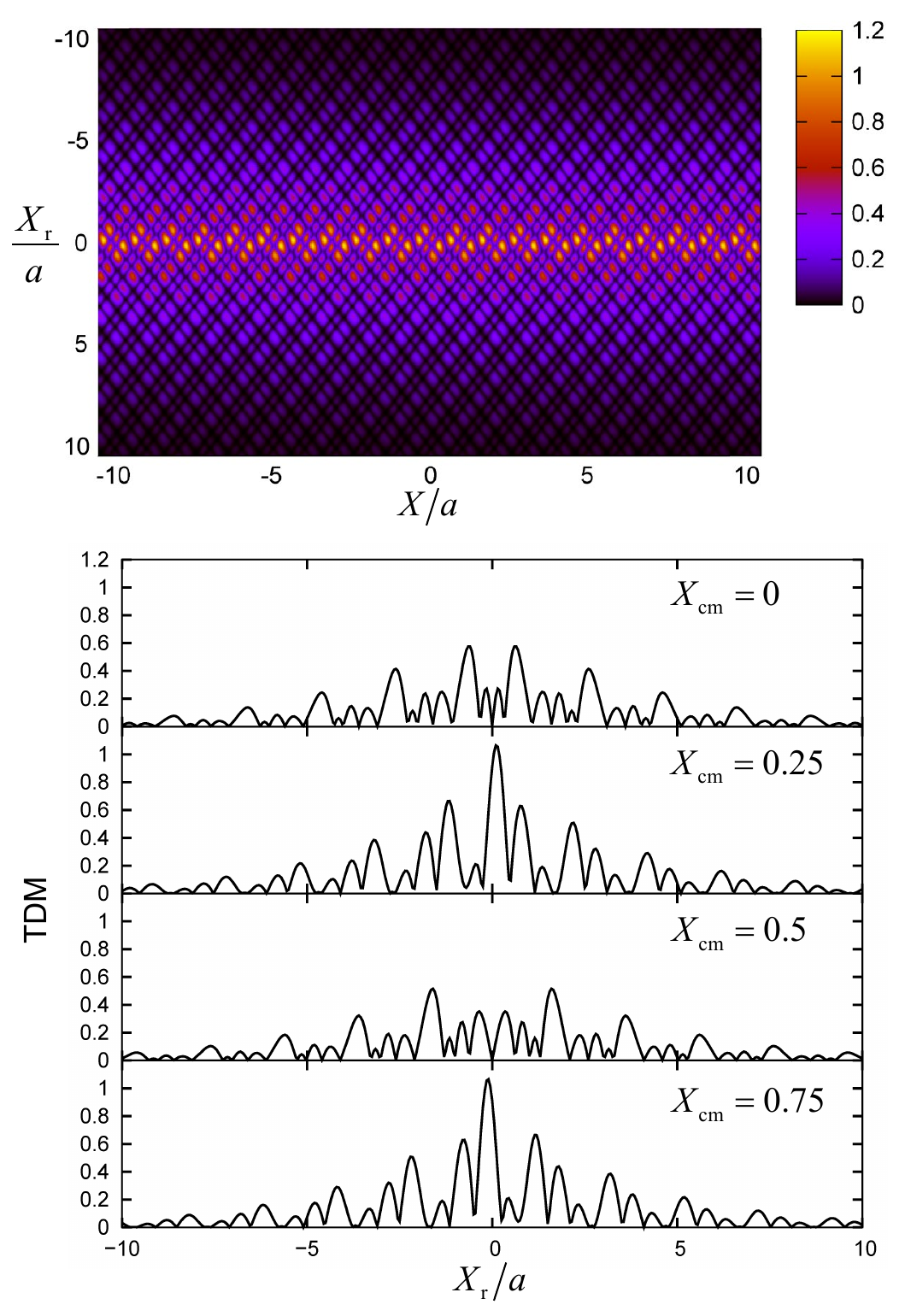}\centering
\caption{Top: frequency-dependent TDM $|\Gamma_s(X,X_{\rm r})|$ of the bound exciton for $\alpha=3$.
Bottom: associated exciton wave function $|\Gamma_s(X_{\rm cm},X_{\rm r}))|$ for various reference positions $X_{\rm cm}$ of the center of mass of the exciton.} \label{fig4}
\end{figure}

Figure \ref{fig4} shows the absolute value of the TDM in an alternative representation, namely, as a function of center-of-mass and relative coordinates of the exciton,
$X = (x+x')/2$ and $X_{\rm r} = x-x'$. The TDM now appears as a broad horizontal stripe, which expresses the translational invariance in the model solid.
The bottom panels show the exciton wave function (in terms of the relative coordinate) for various center-of-mass positions in the unit cell.

For this simple model solid, it is not obvious which of the two representations of the exciton wave function, $\Gamma_s(x,x')$ or $\Gamma_s(X,X_{\rm r})$,
is to be preferred, since both convey similar information. However, we can use the representation $\Gamma_s(X,X_{\rm r})$ to make a comparison with
the basic Wannier model of excitons \cite{Haug}.

According to the 3D Wannier model, the excitons are described using hydrogenic wave functions (as a function of relative coordinate, for arbitrary center-of-mass position).
One finds that the effective Bohr radius of the $1s$ exciton behaves as $a_0 \sim E_b^{-1/2}$, where $E_b = \hbar^2/2m_r a_0^2$ is the exciton binding
energy of the 3D Wannier model ($m_r$ is the reduced electron-hole effective mass).

In 1D, no analytic solution exists for the Wannier model with soft Coulomb interaction. We have obtained numerical solutions of the corresponding
1D hydrogenic Schr\"odinger equation, and we found that one can extract a 1D equivalent, $a_0^{\rm 1D}$, of the Bohr radius as the half-width at half-maximum (HWHM) of the
exciton wave function. It turns out that $a_0^{\rm 1D} \sim E_b^{-\xi}$, where the exponent $\xi$ is close to 0.5.

\subsubsection{1D solid with defects: visualizing localized and charge-transfer excitons}\label{sec:IIIA3}

\begin{figure}
\includegraphics[width=\linewidth]{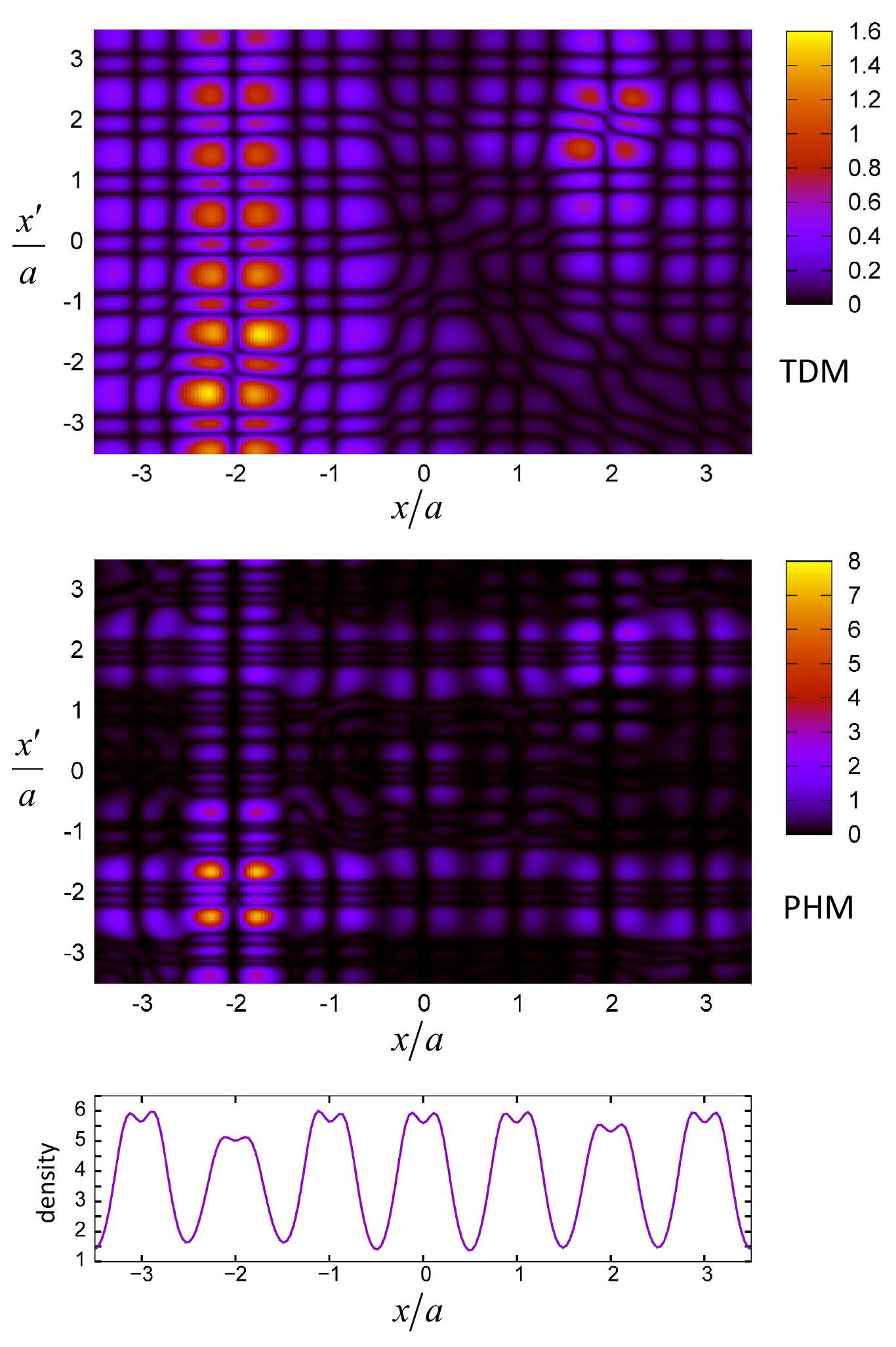}\centering
\caption{Bottom: ground-state density $n_0(x)$ in a 1D supercell consisting of 7 primitive unit cells of the cosine potential (\ref{VKS_cosine}); defects are
simulated by reducing the depth of selected potential wells.
Top and middle: TDM $|\Gamma_s(x,x')|$ and PHM $|\Xi(x,x')|$ of the exciton ($\alpha=2$) within one supercell.
} \label{fig5}
\end{figure}

\begin{figure}
\includegraphics[width=\linewidth]{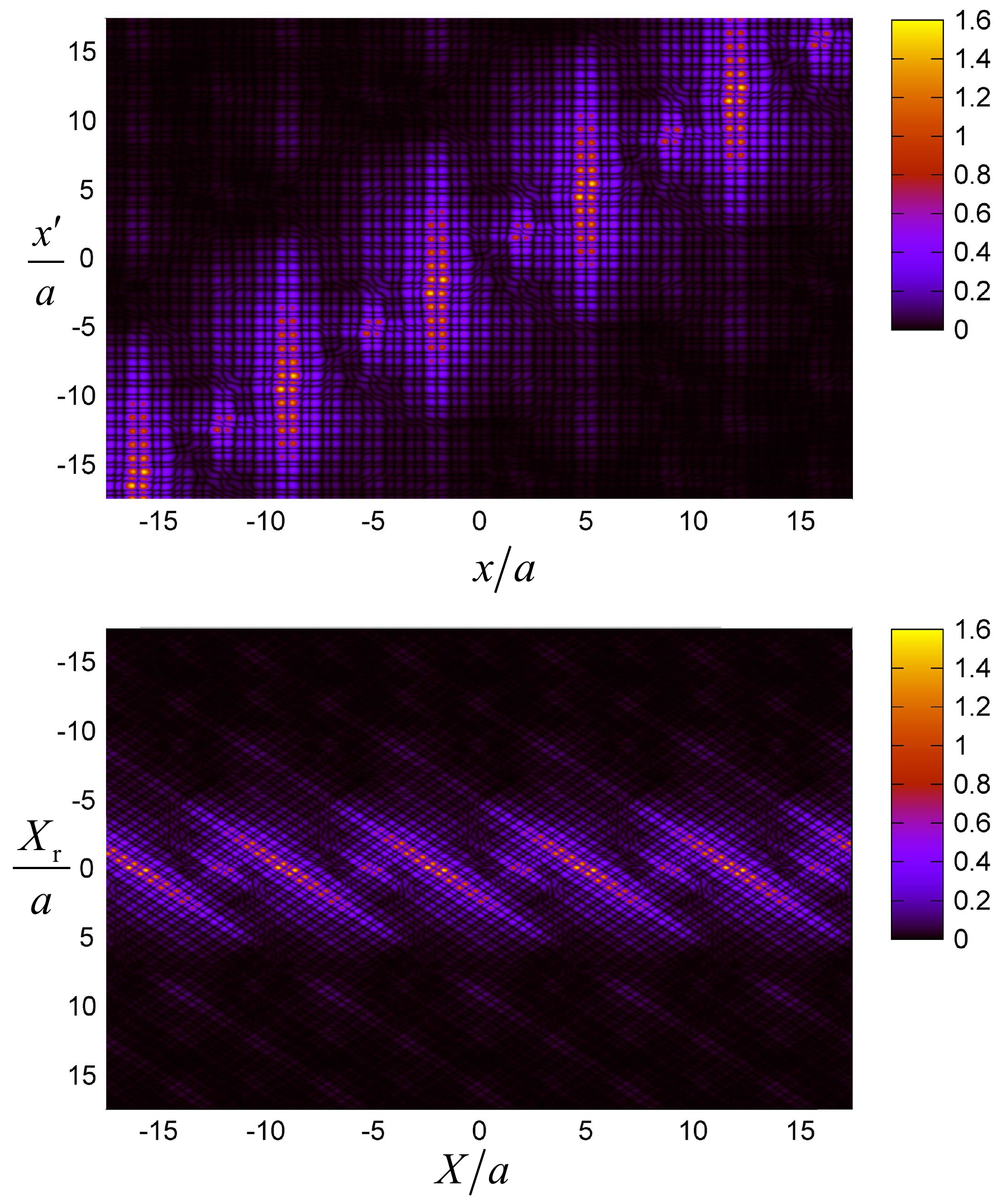}\centering
\caption{Comparison of the two representations of the TDM, $|\Gamma_s(x,x')|$ (top) and  $|\Gamma_s(X,X_{\rm r}))|$ (bottom), for the exciton in the 1D solid
with defects (see Fig. \ref{fig5}).} \label{fig6}
\end{figure}

At equilibrium, the particle-hole map (PHM) offers an alternative tool for visualizing electronic excitations \cite{Li2015,Li2016,Pluhar2018}.
It is defined similarly to the TDM (\ref{TDMns}), as follows:
\begin{eqnarray}\label{PHM}
\Xi^{(n)}_s(\bfr,\bfr') &=& \sum_{ia}[\varphi_i(\bfr')\varphi_a^*(\bfr')|\varphi_i(\bfr)|^2X_{ia}^{(n)} \nonumber\\
&&
{}+ \varphi_i^*(\bfr')\varphi_a(\bfr')|\varphi_i(\bfr)|^2Y_{ia}^{(n)}].
\end{eqnarray}
The PHM provides visual information about the origins and destinations of electrons and holes created during an excitation, and
is therefore particularly suitable to analyze charge-transfer excitation processes in large molecules or molecular complexes \cite{Li2016a}.
Here, we consider the PHM for periodic crystals; in contrast with the TDM, the PHM is itself lattice periodic, and is therefore not very interesting
for excitons in simple solids such as the 1D model treated above.
However, as we will see, the PHM can offer valuable insight into the nature of excitation processes in solids with more complex unit cells.

To give an example,
we consider a supercell consisting of 7 primitive unit cells of the cosine potential, Eq. (\ref{VKS_cosine}). To simulate defects, we modify the amplitude $A$
of the cosine potential in selected cells: specifically, we choose $A=14$ and $A=17$ in cells 2 and 6, respectively, and $A=20$ in all other cells.
The resulting ground-state density is shown in the bottom panel of Fig. \ref{fig5}. The density is slightly reduced at the defect positions.
We then solve the Casida equation as before, using $\alpha=2$. The resulting exciton binding energy, $E_b=0.353$,
is significantly larger than in the absence of the defects ($E_b=0.248$).

In Fig. \ref{fig5} we compare the TDM (top) and the PHM (middle) for the exciton within one supercell. At first glance, both seem to convey similar visual information,
but the physical meaning is different. The vertical streak of the TDM around $x/a=-2$ indicates that the hole strongly localizes at the left defect.
A similar localization of the hole,  but to a lesser degree, occurs at the right defect around $x/a=2$.
The PHM, on the other hand, tells us about the charge-transfer nature of the exciton. The prominent signal at $(x/a,x'/a)=(-2,-2)$ indicates that excitation is mostly
localized at the left defect. However, there are distinct off-diagonal features at $(2,-2)$ and $(-2,2)$, which tell us that there is coherent charge transfer happening
between the two defect sites (in the sense that electrons at one site are associated with holes at the other site, and vice versa).

In Fig. \ref{fig6}, we show the exciton wave functions extending over several supercells, comparing the two representations $\Gamma_s(x,x')$ and $\Gamma_s(X,X_{\rm r})$.
Clearly, in the presence of the defects, the exciton wave function is dramatically changed compared to the pristine case shown in Figs. \ref{fig3} and \ref{fig4},
mainly due to localization. Thus, the example discussed here illustrates that the TDM and PHM can provide complementary information about the
inner mechanisms of excitonic processes.

\subsection{Time-resolved calculations}\label{sec:IIIB}

We now demonstrate that excitonic effects can also be captured in the time domain, via the
time-dependent Kohn-Sham equation in the velocity gauge, as defined in Eq. (\ref{KSA}).
For our 1D model solid, we use the following form of the time-dependent exchange-correlation potential:
\begin{equation}\label{xcpot}
V_{\rm xc}^{\rm LRC}(x,t) = \int dx' f^{\rm LRC}(x,x')\delta n(x',t) \:,
\end{equation}
where $\delta n(x,t) = n(x,t) - n_0(x)$ is the density response. The Hartree potential is ignored here, since it does not contribute to the excitonic binding
(likewise, we ignored the Hartree kernel in the frequency-dependent Casida formalism).
In reciprocal space, one therefore finds $V_{{\rm xc},G}(t) = f^{\rm LRC}_{{\rm xc},GG}\delta n_{\rm G}(t)$, using Eq. (\ref{fxc_G}).
Notice that we set $V_{{\rm xc},G=0}(t)=0$, since the head of $f^{\rm LRC}_{{\rm xc},GG}$ does not contribute to the excitonic interaction in 1D,
as discussed above. This is different in 3D, where the ${\bf G}=0$ contribution is crucial; however, $V_{{\rm xc},{\bf G}=0}(t)$ is ill defined in 3D, which
can be remedied by gauge transforming it into a vector potential \cite{Sun2020b}.

\subsubsection{Resonant versus nonresonant excitation}

We consider time-dependent perturbations of the reciprocal-space form
\begin{equation}\label{A1G}
A_{1,G} = E_0 F(t) \delta_{G,0},
\end{equation}
associated with a uniform short-pulsed field $E(t)=E_0f(t)$ with amplitude $E_0$ and time-dependence
\begin{equation}\label{ft}
f(t) =\sin(\omega_d t) \sin^2 \left(\frac{\omega_d t}{2 N_c}\right)\:, \qquad t \le \frac{2\pi N_c}{\omega_d}\:.
\end{equation}
Here $\omega_d$ is the driving frequency of the external field, $N_c\ge 1$ is the number of cycles contained in the pulse, and $f(t)=0$ for $t> 2\pi N_c/\omega_d$.
With this choice of $f(t)$, the vector potential (\ref{A1G}) vanishes at the end of the pulse, and
the time-dependent TDM becomes gauge invariant [i.e., $\Gamma_s(x,x',t)$ and $\tilde\Gamma_s(x,x',t)$ coincide]
once the system reaches the stage of free time propagation.

\begin{figure}
\includegraphics[width=\linewidth]{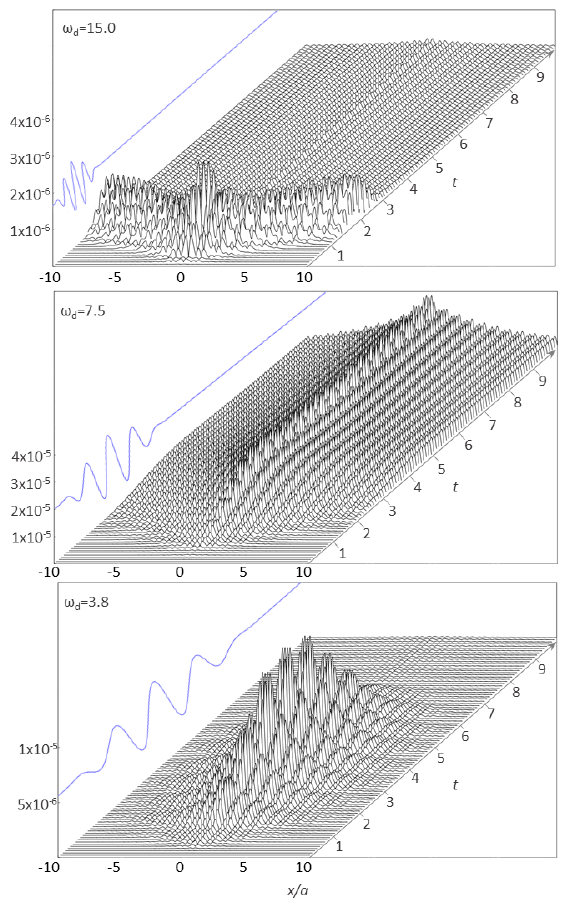}\centering
\caption{Time-dependent TDM for the 1D solid of Fig. \ref{fig1}, subject to 5-cycle pulsed fields of the form given in Eq. (\ref{ft}), with $E_0=0.0001$,
for three different values of $\omega_d$: above resonance (top), on resonance (middle) and below resonance (bottom) with the 1D exciton (here, $\alpha=2$). } \label{fig7}
\end{figure}

Figure \ref{fig7} shows waterfall plots of the time-dependent TDM (using $\alpha=2$) in three cases: for above-resonant ($\omega_d=15.0$), resonant ($\omega_d=7.5$) and
below-resonant ($\omega_d=3.8$) excitation. The electric-field amplitude is the same in each case, $E_0=0.0001$, corresponding to a very weak excitation.
The three cases are strikingly different.  Above and below resonance, the system shows a pronounced response of the TDM as long as the pulse is present, but
after the end of the pulse the signal essentially disappears. In the above-resonant case, the excitation energy is much greater than the band gap, thus
promoting the carriers well into the conduction band and into the incoherent single-particle regime, far from the exciton.
On the other hand, the below-resonant response can be viewed as a transient, quasi-static polarization effect.
Only the resonant excitation leads to a time-dependent TDM that
persists after the pulse is over, leading to a time-dependent TDM that essentially maintains its shape, apart from some minor oscillations at the frequency of the exciton.

\begin{figure}
\includegraphics[width=\linewidth]{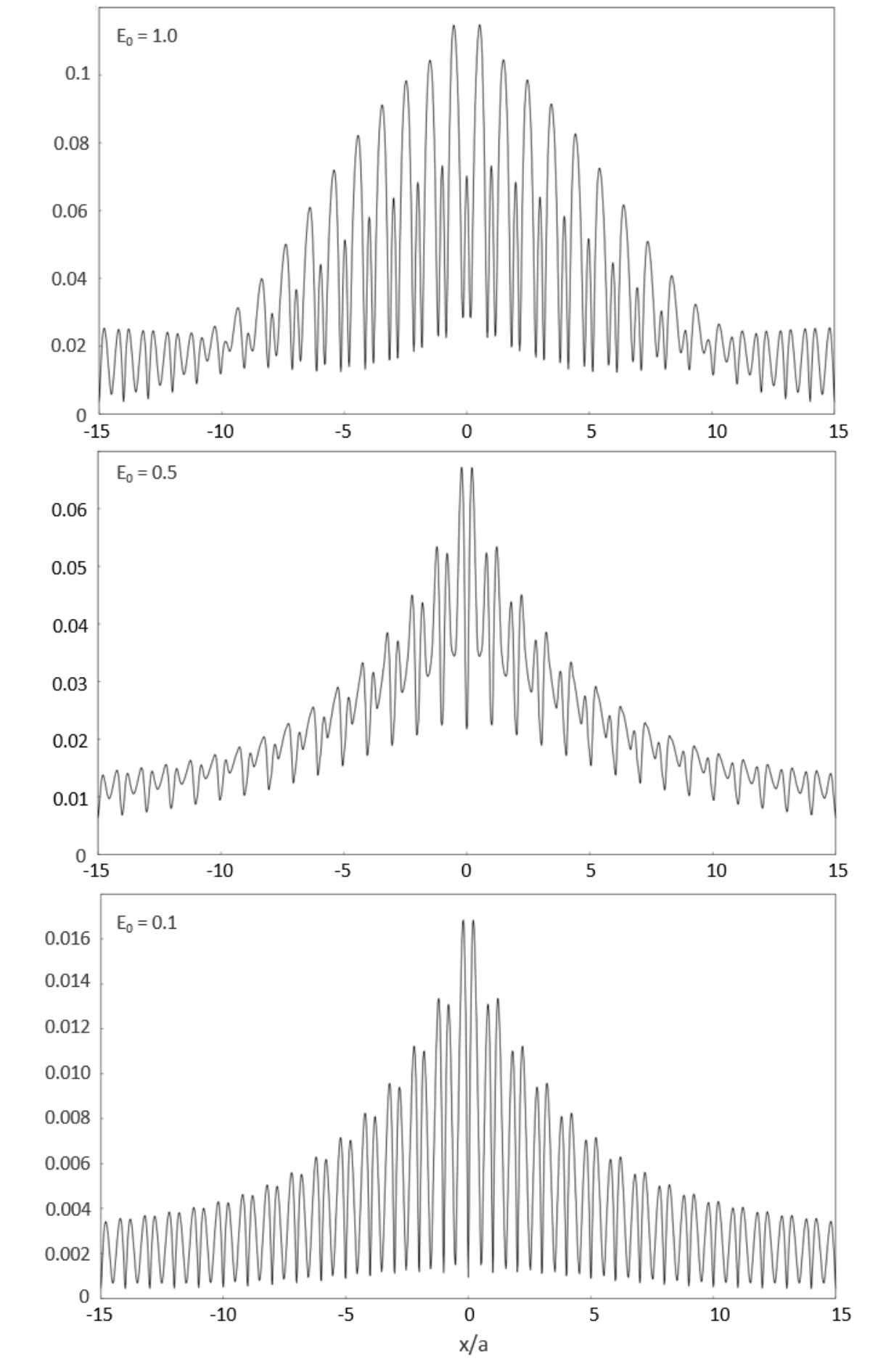}\centering
\caption{Time average of the time-dependent TDM for resonant excitation (see middle panel of Fig \ref{fig7}), calculated after the end of the pulse.} \label{fig8}
\end{figure}

\subsubsection{Strong excitations}

One would expect that the results obtained from real-time propagation are consistent with frequency-dependent linear response.
To demonstrate that this is indeed the case, we calculate the time average of the resonant time-dependent TDM over a time interval of 10 a.u. after the end of the
pulse. Indeed, we find an almost perfect agreement with linear response for field strengths up until $E_0=0.1$:
in this case, the averaged time-dependent TDM, shown in the bottom panel of Fig. \ref{fig8}, is indistinguishable from the frequency-dependent TDM.

However, differences start to develop for stronger excitations, see the middle panel ($E_0=0.5$) and top panel ($E_0=1.0$) of Fig. \ref{fig8}.
Clearly, for $E_0=1.0$ the exciton wave function no longer has an exponential envelope, but has become significantly distorted and broadened.
This suggests that nonlinear effects start to become noticeable once the peak field strength of the laser pulse exceeds 0.1.
Of course, entering the nonlinear regime for $E_0 \gtrsim 0.1$ also means that our expression for the xc potential, Eq. (\ref{xcpot}), is no longer
formally justified \footnote{On the other hand, this does not necessarily mean that the xc potential (\ref{xcpot}) will perform poorly beyond the weakly perturbed regime,
only that one should  proceed with caution.
There are many examples in (TD)DFT where approximations are used very successfully in situations that are formally not well justified,
most notably the adiabatic local-density approximation (ALDA).}.

\begin{figure}
\includegraphics[width=\linewidth]{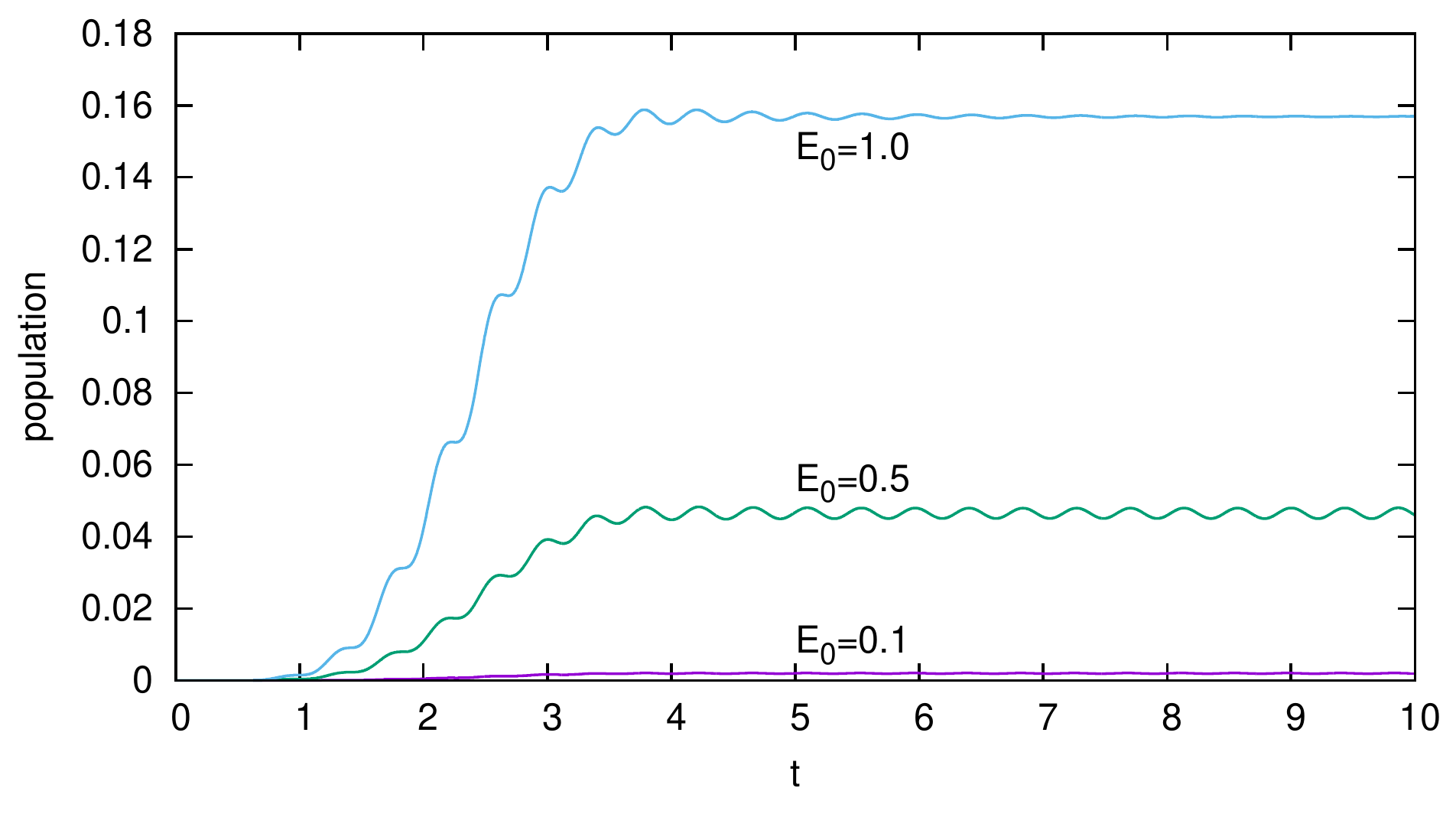}\centering

\includegraphics[width=\linewidth]{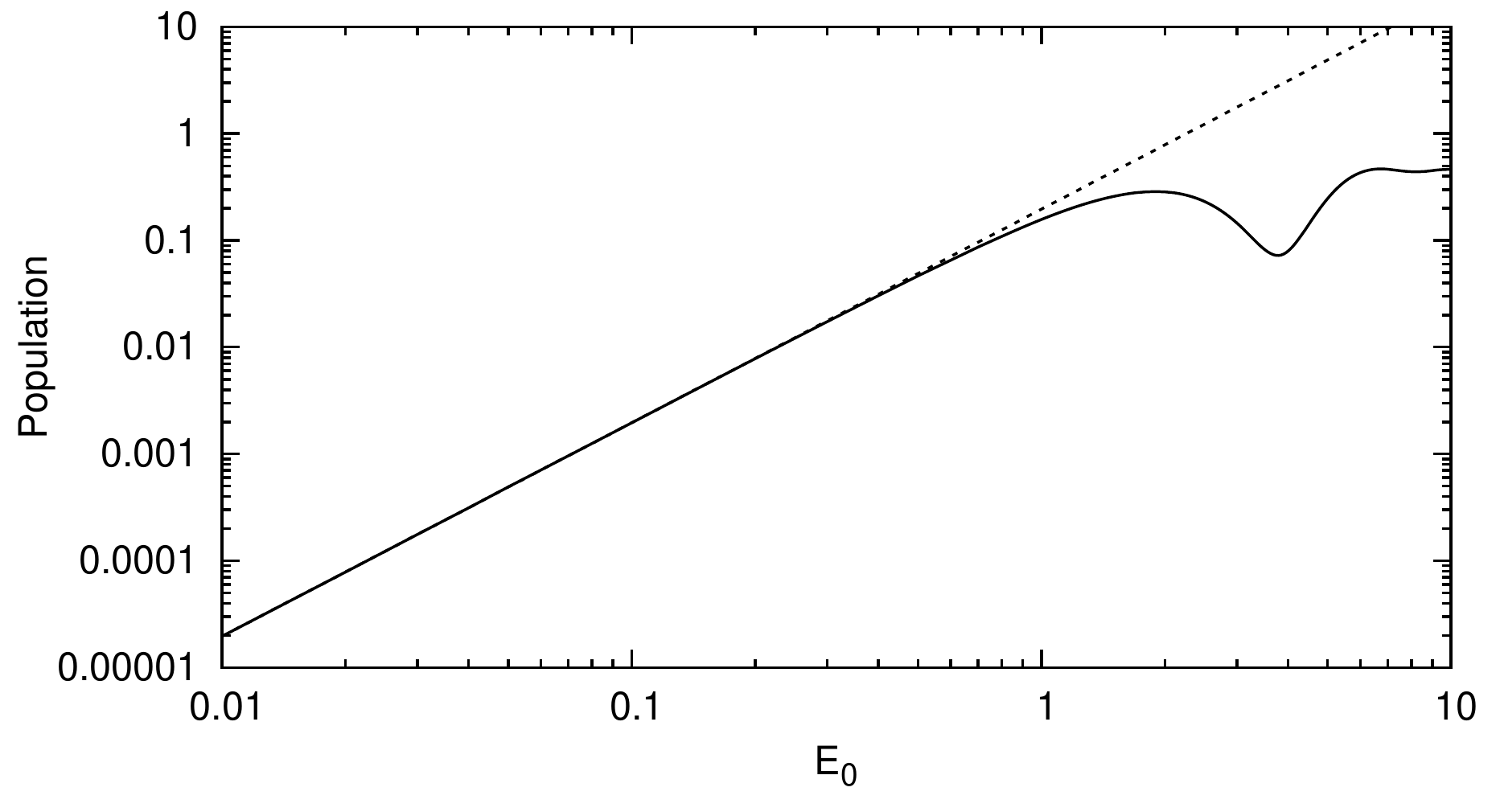}
\caption{Top: time dependence of the excited-state population for the three cases shown in Fig. \ref{fig8} (resonant excitation with $\omega_d=7.5$).
Bottom: average excited-state population at the end of the pulse, as a function of $E_0$. The dashed line indicates a behavior $\sim E_0^2$. } \label{fig9}
\end{figure}

To analyze this further, we plot in Fig. \ref{fig9} the population $P_{\rm ex}(t)$ of the initially empty bands, i.e., the number of excited (Kohn-Sham) electrons
per unit cell,
promoted into the initially empty bands marked in red in Fig. \ref{fig1}. As shown in the top panel,
the excited-state population $P_{\rm ex}(t)$ rises sharply during the pulse and then stabilizes, apart from some small oscillations.
Denoting the time average after the end of the
pulse by $\bar P_{\rm ex}$, we find $\bar P_{\rm ex} = 0.00196$, 0.0465 and 0.157 for $E_0=0.1$, 0.5 and 1, respectively.

The bottom panel of Fig. \ref{fig9} shows $\bar P_{\rm ex}$ as a function of $E_0$ on a logarithmic scale.
The straight-line behavior for small $E_0$ indicates that, as expected for a resonant one-photon excitation process,
the number of excited electrons grows quadratically with $E_0$, i.e., proportional to the peak intensity of the pulse.
As $E_0$ approaches 1, $\bar P_{\rm ex}$ starts to fall behind the quadratic behavior. Again, this is not unexpected: as the population of the bands
changes significantly, the energy levels shift and the laser field detunes; this makes the excitation process less effective.
In TDDFT, this detuning effect is a well-known weakness of adiabatic approximations to the time-dependent xc potential, leading, among other things,
to inability to describe Rabi oscillations \cite{Maitra2016,Luo2016}.

Thus, we conclude that the time-dependent exciton wave function reflects the transition from the linear to the nonlinear regime in the form of an
increasing peak height with a gradual
change of shape, although a bound exciton remains recognizable well beyond the linear regime. For extremely strong excitations, outside the range of
validity of the present approach, the exciton wave function will distort more and more strongly, and will eventually dissolve.

\subsubsection{Exciton dissociation in static electric fields}

In the presence of a uniform static electric field, the bound states of a hydrogenic Hamiltonian become metastable: even a weak field leads to a suppression of the
Coulomb potential, which allows the electron to tunnel out \cite{Oppenheimer1928}. The field-induced dissociation of excitons, which is in many ways similar
to the tunneling ionization of the H-atom,
has been widely studied both theoretically and experimentally \cite{Haastrup2016,Meng2017,Heckotter2018,Massicotte2018,Kamban2019}.
In practice, one is interested in the rates at which the excitons dissociate, and how these rates depend on the material.

The atomic unit of electric field strength (the field experienced by an electron at a distance of one Bohr radius $a_0$ from a proton)
is ${\cal E}_0 = e^2/(4\pi \epsilon_0 a_0^2) = 5.14\times 10^{11}$ V/m. An applied field whose strength approaches ${\cal E}_0$
causes an H-atom to ionize. For free carriers
in materials with effective mass $m^*=\mu m$ and effective charge $e^* = e/\sqrt{\epsilon_r}$, the atomic unit of electric field becomes
${\cal E}_0^* = (\mu^2/\epsilon_r^3){\cal E}_0$, which can be less than ${\cal E}_0$ by several orders of magnitude.
Accordingly, typical field strengths at which exciton dissociation becomes noticeable range anywhere from $10^6 - 10^8$ V/m,
depending on the material \cite{Haastrup2016,Meng2017,Heckotter2018,Massicotte2018,Kamban2019}.

We now want to find out how a static electric field influences the exciton wave function, and whether we can observe signatures of dissociation. For this purpose,
we include an additional term into the time-dependent vector potential  (\ref{A1G}) and write
\begin{equation}
A_{1,G} = [E_0 F(t) + E_{\rm stat} t] \delta_{G,0} \:.
\end{equation}
In other words, together with the laser pulse of peak strength $E_0$ that creates the exciton,
we switch on a uniform static field of strength $E_{\rm stat}$ at time $t=0$.

\begin{figure*}[t]
\includegraphics[width=\linewidth]{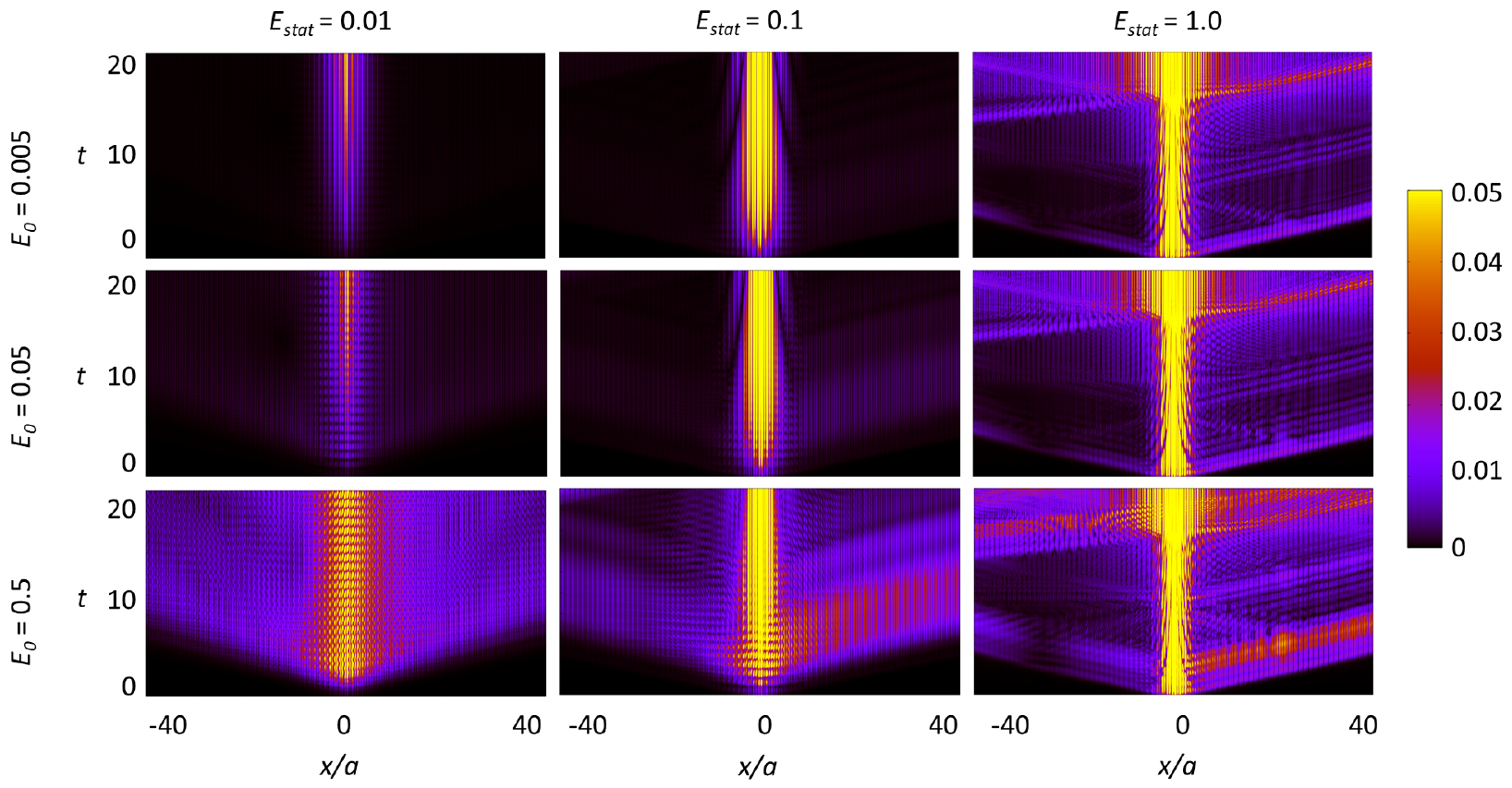}
\caption{Time dependent exciton wave functions, following resonant excitation with pulses of strength $E_0$, subject to
static electric fields $E_{\rm stat}$. The exciton wave functions are plotted as a function of electronic coordinate $x'$ for the hole fixed at $x=0$;
the peak height is capped at 0.05. The wave function develops an asymmetry under sufficiently strong static fields, indicating dissociation.} \label{fig10}
\end{figure*}

In Fig. \ref{fig10} we illustrate the time-dependent exciton wave function following resonant excitation with pulses of strength $E_0=0.005$, $0.05$ and $0.5$, subject to
static electric fields of strength $E_{\rm stat}=0.01$, $0.1$ and 1.
A static field has important consequences for Bloch electrons in materials with a gap \cite{Kittel}:
it causes interband transitions  via Zener tunneling, and it affects the carrier dynamics by causing Bloch oscillations and Wannier-Stark ladders
\cite{Krieger1986,Kruchinin2018}. Since our focus is on coherent electron-hole pairs, Bloch oscillations are not visible here.
On the other hand, the tunneling effect is clearly seen in our calculations: the field-induced interband transitions
lead to a steady increase of excited-state population. As a result, the central peak of the exciton wave function keeps increasing
in height; in Fig. \ref{fig10}, we have capped the peak height at a value of 0.05, because this would otherwise overshadow the main effect.

We observe that for sufficiently strong fields, of order $E_{\rm stat}=0.1$ or higher, the exciton wave function develops
an asymmetry following the initial pulse which triggers the exciton. The exciton wave function here represents the distribution of an electron around a hole fixed at $x=0$.
Thus, we can clearly observe the flux of the electron moving to the right.
Integrating over the outgoing flux can provide a practically useful measure of the rate of exciton dissociation.
Here, the rate is small compared to the rate of interband tunneling. This is likely a consequence of the extremely large
oscillator strength of the bound exciton in our 1D model system.

In the last panel ($E_0=0.5$ and $E_{\rm stat}=1$), we observe another interesting effect:
it can be clearly seen that the flux of the outgoing electron re-enters from the left after some time. In other words,
the system behaves as a ring of circumference $N_k a$, where $N_k$ is the number of $k$-points (here, $N_k=200$).
Choosing $N_k$ sufficiently large is thus essential in order to avoid unphysical finite-size effects in simulating exciton dynamics.


\section{Conclusion}\label{sec:IV}

In this paper, we have shown how excitons can be visualized using Kohn-Sham TDDFT. The method, based on the single-particle
TDM, can be applied in frequency-dependent linear-response as well as in the real-time regime.
In the appropriate limit of weak perturbations, the frequency-dependent linear-response and the real-time versions of
the TDM lead to the same representation of the exciton wave function. However, the real-time TDM can be extended beyond
the linear regime, giving access to the description of exciton dynamics under strong, ultrafast excitations.

We have illustrated the features and capabilities of the Kohn-Sham TDM for 1D model solids in various scenarios.
As the example of charge-transfer excitons in a solid with defects shows, the TDM (together with its cousin, the PHM)
delivers a spatially resolved exciton wave function which allows one to extract useful information about the excitation mechanism in the material.

In the real-time domain, we have discussed the formation and dynamics of excitons following short-pulsed excitations.
The time-dependent exciton wave function exhibits very different behavior depending on whether the excitation is resonant or
nonresonant. As the excitation strength is increased and more and more carriers are promoted across the gap, the exciton wave function begins to
distort, but the essential features of a bound exciton are preserved well into the nonlinear regime.
In the presence of static electric fields, the exciton wave function displays signatures of dissociation.

We have thus shown that the Kohn-Sham TDM is a versatile and powerful visualization tool for excitons.
The time-dependent exciton wave function introduced in this paper is computationally easy to implement
in conjunction with time-dependent Kohn-Sham calculations for extended systems such as periodic solids, nanostructures, or large molecules.
However, the TDM can provide more than just visual information, and could in fact be used in various ways for a more
quantitative analysis of exciton dynamics, for instance to extract dissociation rates under the influence of
a bias, or charge separation rates at interfaces. Given the increasing use and availability of real-time electronic structure approaches in chemistry and
materials science \cite{Li2020}, this may open up many new and promising applications in excitonics.

\acknowledgments{This work was supported by NSF grant No. DMR-1810922. }


\bibliography{TDM_refs}

\end{document}